\documentclass[aps,amsfonts,amsmath,prd,preprint,nofootinbib]{revtex4}
\newcommand{\beq}{\begin{equation}}
\newcommand{\eeq}{\end{equation}}

\usepackage{epsfig,bbm,cancel}
\usepackage[breaklinks=true]{hyperref}

\begin{document}

\title{On DBI Textures with Generalized Hopf Fibration}

\author{Handhika S. Ramadhan}
\email{ramadhan@teori.fisika.lipi.go.id}
\affiliation{Group for Theoretical and Computational Physics\\ Indonesian Institute of Sciences\\ Kompleks Puspiptek Serpong Tangerang 15310, Indonesia\\ and\\ Department of Physics University of Indonesia, Kampus UI Depok 16424, Indonesia. }

\def\changenote#1{\footnote{\bf #1}}

\begin{abstract}
In this letter we show numerical existence of $O(4)$ Dirac-Born-Infeld (DBI) Textures living in $(N+1)$ dimensional spacetime. These defects are characterized by $S^{N}\rightarrow S^{3}$ mapping, generalizing the well-known Hopf fibration into $\pi_{N}(S^{3})$, for all $N>3$. The nonlinear nature of DBI kinetic term provides stability against size perturbation and thus renders the defects having natural scale.
\end{abstract}

\maketitle
\thispagestyle{empty}
\setcounter{page}{1}

Topological defects are static solutions (solitons) of classical field equations whose existence is due to the topology of their boundary conditions~\cite{vilenkinshellard, kibble, rajaraman, mantonsutclife}. {\it Domain walls}, {\it cosmic strings}, {\it monopoles}, and {\it textures} are examples of defects characterized by the nontrivial homotopy groups of their vacuum manifold $\mathcal{M}$, $\pi_{N}(\mathcal{M})\neq\mathbb{I}$ with $N=0,1,2,3$ correspondingly. Among other defects, textures are rather different in the sense that the field is everywhere in the vacuum manifold~\cite{davis, turok}. The nontrivial homotopy group is achieved by mapping from coordinate space into the field space. In $(3+1)$ dimensions the simplest example is an O(4) texture spontaneously broken to O(3), with vacuum manifold $O(4)/O(3)\cong S^{3}$. At infinity there is a one-to-one mapping\footnote{The spatial dimensions have topology $\mathbb{R}^{3}$, but the boundary conditions, that the field should approach the same value at infinity, compactifies it into a $3$-sphere, $\mathbb{R}^{3}\cup\{\pm\infty\}\cong S^{3}$~\cite{mantonsutclife}.} $S^{3}\rightarrow S^{3}$, with relevant homotopy group $\pi_{3}(S^{3})=\mathbb{Z}$. 

Topology, however, provides only a necessary, yet not sufficient, condition for the stability of defects. Derrick's theorem~\cite{derrick} shows that textures are unstable. One way to stabilize it is by having higher order kinetic terms, as shown by Skyrme~\cite{skyrme}. More recently scalar field theories with noncanonical kinetic terms have been extensively studied in~\cite{noncanon, chiraldbi} where it is shown that stable defects exist. In~\cite{dbipaper} we prove numerically that $SO(N)$ textures with Dirac-Born-Infeld (DBI) kinetic term can be stabilized in any arbitrary spatial dimensions $N\geq 3$. Stability is achieved by the nonlinear nature of DBI form, which can be Taylor expanded at any desired order to give sufficient higher terms to evade Derrick's theorem. Mathematically, these defects fall into homotopy class $\pi_{N}(S^N)$, which is nontrivial. Physically, they are examples of {\it extended} defects (or solitons), {\it i.e.}, stable defects living in higher dimensions.  

We further notice that generalization of the group symmetry to $O(N+1)$ might not be necessary; {\it i.e.}, that $O(4)$ stable textures can exist in any arbitrary codimensions ($>3$). Here we realize the analogy with the Hopf solitons (Hopfions)~\cite{faddeev, afz}, where $O(3)$ textures lives in three spatial dimensions, resulting in the $S^{3}\rightarrow S^{2}$ mapping. They belong to the category of Hopf fibration (see, for example,~\cite{hopf}), with nontrivial homotopy group $\pi_{3}(S^{2})\neq\mathbb{I}$. Topological solitons with higher homotopy Hopf fibration have also been investigated in~\cite{exotic}. It is the purpose of this letter to show that $O(4)$ solitonic textures with generalized Hopf fibration can exist. Here we limit ourselves only to spherically symmetric ansatz and topological charge unity.  
 
The model we discuss has the following Lagrangian~\cite{dbipaper}
\begin{equation}
\label{eq:chiraldbisigma}
{\cal{L}}_{DBI}=\beta^{2}_{N}\left[\sqrt{1+\frac{1}{\beta^{2}_{N}}\partial_{\mu}\Phi^{i}\partial^{\mu}\Phi^{i}}-1\right].
\end{equation} 
The mass-scale parameter, $\beta_{N}$, is a dimensionful parameter with dimension $\beta_{N}^{2}\sim[M]^{(N+1)}$. Under a change of variables to the dimensionless units
\begin{equation}
x\rightarrow\frac{x}{M},\  \ \phi\rightarrow M^{\frac{N-1}{2}}\phi,\ \ \beta_{N}\rightarrow M^{\frac{N+1}{2}}\beta,
\end{equation}
the rescaled parameter $\beta$ (as well as the Lagrangian) becomes dimensionless. It controls the strength of nonlinear DBI coupling. For large $\beta$, the Lagrangian reduces to
\begin{equation}
{\cal{L}}\sim\frac{1}{2}\partial_{\mu}\Phi^{i}\partial^{\mu}\Phi^{i},
\end{equation}
the ordinary nonlinear sigma model. We can construct a (topological) current
\begin{equation}
\sqrt{-g}j^{\mu}=\frac{1}{12\pi^{2}}\epsilon^{\mu\alpha_{1}\alpha_{2}\ldots\alpha_{N-1}}\epsilon_{a_{1}a_{2}\ldots a_{N}}
\Phi^{a_{1}}\partial_{\alpha_{1}}\Phi^{a_{2}}\partial_{\alpha_{2}}\Phi^{a_{3}}\ldots\partial_{\alpha_{N-1}}\Phi^{a_{N}},
\end{equation}
which is trivially conserved and whose charge is nonzero integers.

In~\cite{dbipaper} we show that the static energy~\footnote{Without loss of generality we can set $\beta=1$ for simplicity.} 
\begin{equation}
\label{eq:energy}
E=\int\left(1-\sqrt{1-E_{2}}\right)d^{N}x,
\end{equation}
with $N$ the number of spatial dimensions and $0<E_{2}\equiv\partial_{i}\phi_{a}\partial^{i}\phi_{a}<1$ gives, under Derrick's theorem, the following conditions
\begin{eqnarray}
\label{eq:rescaledconditions}
\frac{E_{2}}{\sqrt{1-E_{2}}}-N(1-\sqrt{1-E_{2}})=0,\nonumber\\
\frac{E_{2}}{(1-E_{2})^{3/2}}+\frac{(1-2N)E_{2}}{\sqrt{1-E_{2}}}+N(N+1)(1-\sqrt{1-E_{2}})\geq0.
\end{eqnarray}
It is easy to see that for each $N\geq 3$ there always exists a nonzero positive value of $E_{2}$ satisfying the above conditions. In other words, stable extended solitons can exist in higher codimensions\footnote{The case for $N=3$ is discussed in~\cite{chiraldbi}.}.

Unlike the hedgehog ansatz in~\cite{dbipaper} which enjoys $O(N)$ symmetry, here we require only $O(4)$ invariant ansatz,
\begin{equation}
\Phi_{i}=(\cos\alpha(r),\sin\alpha(r)\cos\theta,
\sin\alpha(r)\sin\theta\cos\varphi,\sin\alpha(r)\sin\theta\sin\varphi).
\end{equation}
This yields the following Lagrangian
\begin{equation}
\label{eq:lagchiral}
{\cal{L}}=\beta^{2}\left[\sqrt{1-K}-1\right],
\end{equation}
with
\begin{equation}
\label{eq:K}
K\equiv\frac{1}{\beta^{2}}\left(\alpha'^{2}+\frac{2\sin^{2}\alpha}{r^{2}}\right),
\end{equation} 
and `primes' denoting derivative with respect to $r$. The field equation is
\begin{equation}
\label{eq:dbiskyrme}
\left[\frac{\alpha' r^{N-1}}{\sqrt{1-K}}\right]'
-\frac{\sin 2\alpha}{2r^{3-N}\sqrt{1-K}}=0.
\end{equation}
To obtain topological solitonic solutions the appropriate boundary conditions should be
\begin{eqnarray}
\label{eq:bound1}
\alpha(0)=m\pi,\\
\label{eq:bound2}
\alpha(\infty)\rightarrow 0,
\end{eqnarray}
with $m$ positive integers, corresponding to the winding number (topological charge) of the solutions. Conditions (\ref{eq:bound1})-(\ref{eq:bound2}) are essential for the nontrivial winding number\footnote{In this letter we only consider topological charge unity, so we set $m=1$.}. The boundary at $r=0$, condition (\ref{eq:bound1}), ensures the nonsingularity of solitons. It is easy to see at $r=0$ that unless  $\alpha(0)=m\pi$, with\footnote{$m=$half-integers results in the fractional topological charge, which signals topological instability.} $m$ integers, then Eq.(\ref{eq:dbiskyrme}) blows up. The second boundary, (\ref{eq:bound2}), is needed for the finiteness of energy. For static field the energy density is just the zeroth component of the energy-momentum tensor, {\it i.e.},
\begin{eqnarray}
\label{eq:energy}
E_{N}&=&\int d^{N}x\ \varepsilon,\nonumber\\
&=&\Omega_{N-1}\beta^{2}\int^{\infty}_{0}\left[1-\sqrt{1-K}\right]r^{N-1}dr,
\end{eqnarray} 
where $\Omega_{N-1}$ is the hypersurface area of an $(N-1)$-sphere. It is then trivial to see that unless $\alpha(r)\rightarrow 0$ at $r\rightarrow\infty$ then the functional $K$ does not vanish, which implies that the integrand is not zero. This results in the divergence of the static energy.

To have regularity at $r=0$, the expansion of $\alpha(r)$ around the origin should take the form of positive semi-definite power law as the most general expansion,
\begin{equation}
\label{eq:expansion}
\alpha(r)=\pi+\alpha_{1}\ r+\alpha_{3}\ r^{3}+O(r^{5}).
\end{equation}
Note that only odd powers survive due to the self-consistence of this expansion with the field equation, Eq.(\ref{eq:dbiskyrme}). Here $\alpha_{1}$ is a coefficient that cannot be determined from the expansion, while the the next to leading coefficients ({\it e.g.} $\alpha_{3}$, $\alpha_{5}$, $\cdots$) for each codimension $N$ are completely determined by $\alpha_{1}$ and $\beta$. For instance, for $N=4$ we have, to the lowest order,
\begin{equation}
\alpha_{3}=\frac{\alpha_{1}^{3}\ (3-10\ \alpha_{1}^{2})}{3\ (6\ \beta^{2}-13\ \alpha_{1}^{2})}.
\end{equation}
Hence we see that the linear term, $O(r)$, is crucial for the nontriviality of the higher order coefficients. We solve the equation numerically by means of shooting method, treating the undetermined coefficient $\alpha_{1}$ as a shooting parameter. 

\begin{figure}[htbp]
\centering\leavevmode
\epsfysize=9cm \epsfbox{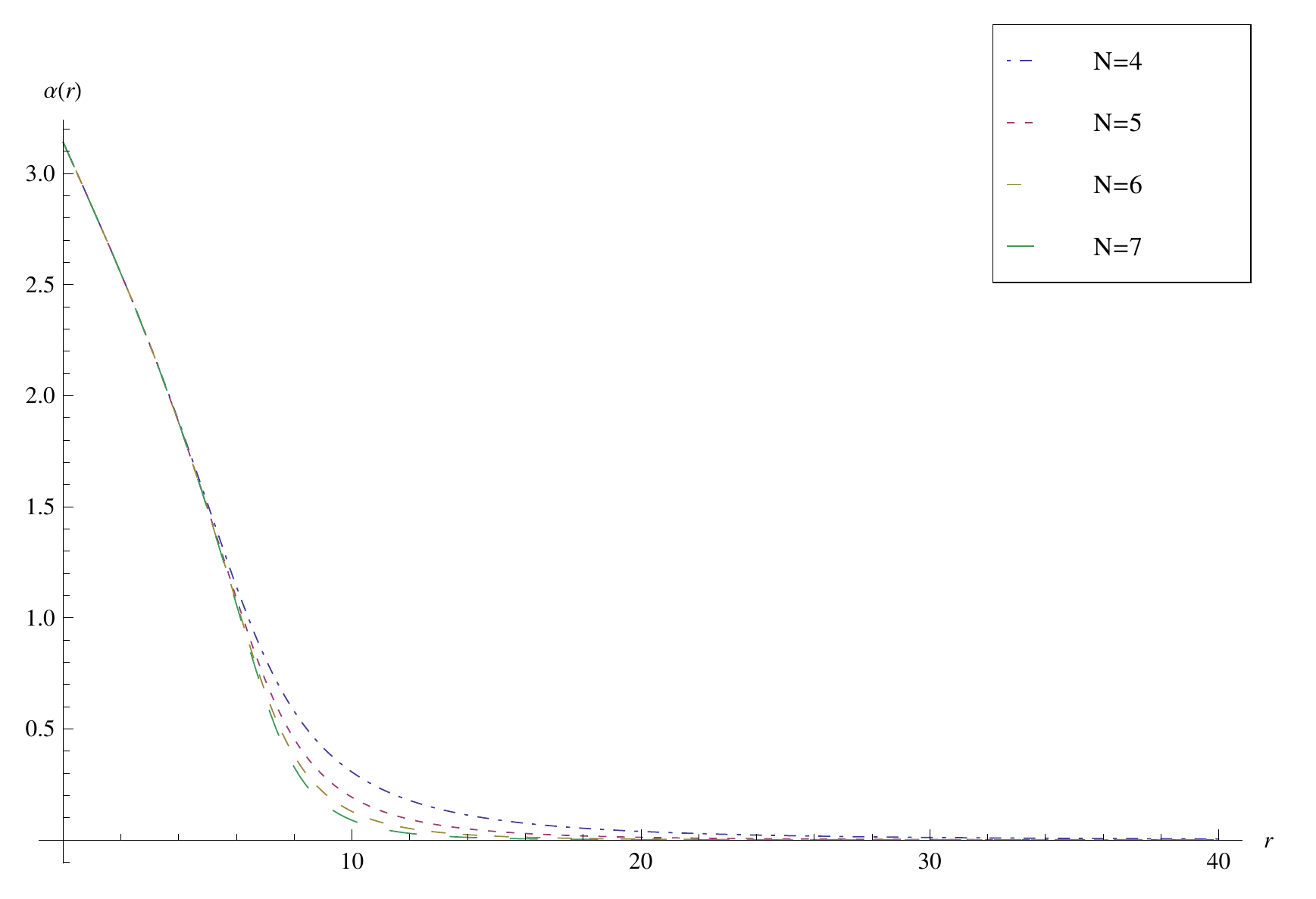}
\caption {$O(4)$ DBI Textures (with $\beta=1/2$) in various codimensions.}
\label{hopfdim}
\end{figure}

As mentioned above, parameter $\beta$ also represents the nonlinearity degree of the theory. Our DBI texture is weakly coupled for $\beta>1$, and strongly coupled for $0<\beta\leq 1$. In the strongly coupled regime we can no longer ignore the higher order expansion of Lagrangian (\ref{eq:chiraldbisigma}); {\it i.e.}, we must consider the full nonlinear field equation.

In Fig.~\ref{hopfdim} we show solutions for several codimensions. Around the core all solutions possess the same gradient field ($\alpha_{1}$), and they start to deviate at some finite distance, the higher codimensions drop faster than the lower ones. As we argued in~\cite{dbipaper}, in the asymptotic regime Eq.(\ref{eq:dbiskyrme}) can be solved analytically with the nontrivial power-law solution given by
\begin{equation}
\alpha(r)\approx\frac{b}{r^{N-1}},
\end{equation}
where $b$ is some undetermined constant of integration. Thus, as the codimension $N$ increases the inverse-power-law solutions get higher inverse power and as a result they fall of faster at large distance. 

\begin{figure}[htbp]
\centering\leavevmode
\epsfysize=9cm \epsfbox{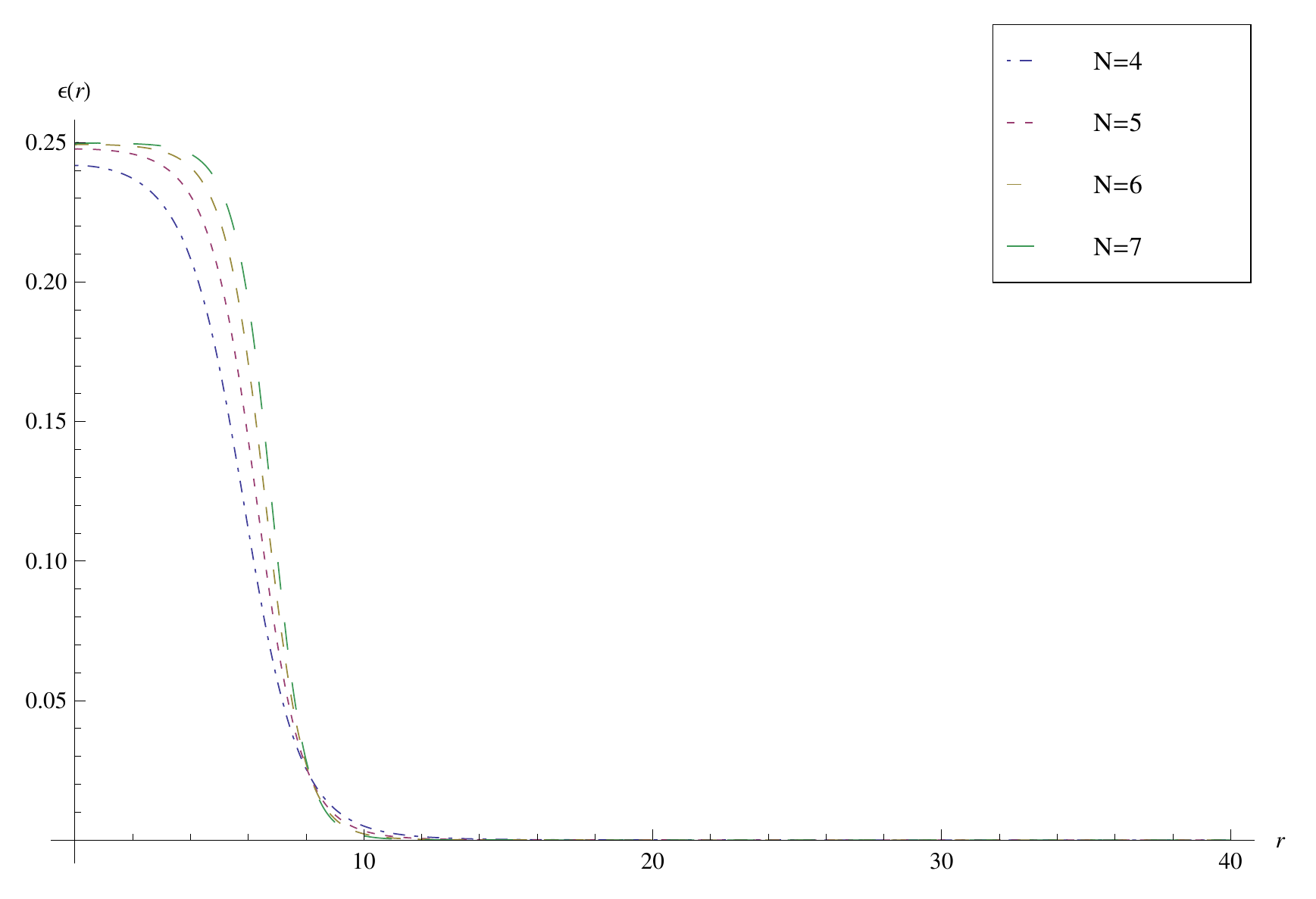}
\caption {Energy densities of $O(4)$ DBI Textures (with $\beta=1/2$) in various codimensions.}
\label{energyden}
\end{figure}

We also present the resulting energy densities for solitonic solutions with various codimensions in Fig.~\ref{energyden}. Their profiles are finite at the cores while at large distance they fall fast enough\footnote{Note that despite the energy density is very much localized the asymptotic behavior is not exponential, but rather power-law, fall off.}. The regularity at the cores is due to the form of expansion~(\ref{eq:expansion}), which can be seen by expanding the energy densities, $\varepsilon(r)$, around the origin ($r=0$)
\begin{equation}
\varepsilon(r)\approx\left(\beta^{2}-\beta^{2}\sqrt{1-\frac{3\alpha_{1}^{2}}{\beta^{2}}}\right)+\frac{\alpha_{1}(15\ \alpha_{3}-\alpha_{1}^{3})}{3\sqrt{\frac{\beta^{2}-3 \alpha_{1}^{2}}{\beta^{2}}}}\ r^{2}+O(r^{4}).
\end{equation}
This behavior is essential for the finiteness of the total static energy. It is also revealed that the energy densities of the higher $N$-solitons fall even faster, thus becoming more localized in space.

\begin{figure}[htbp]
\centering\leavevmode
\epsfysize=9cm \epsfbox{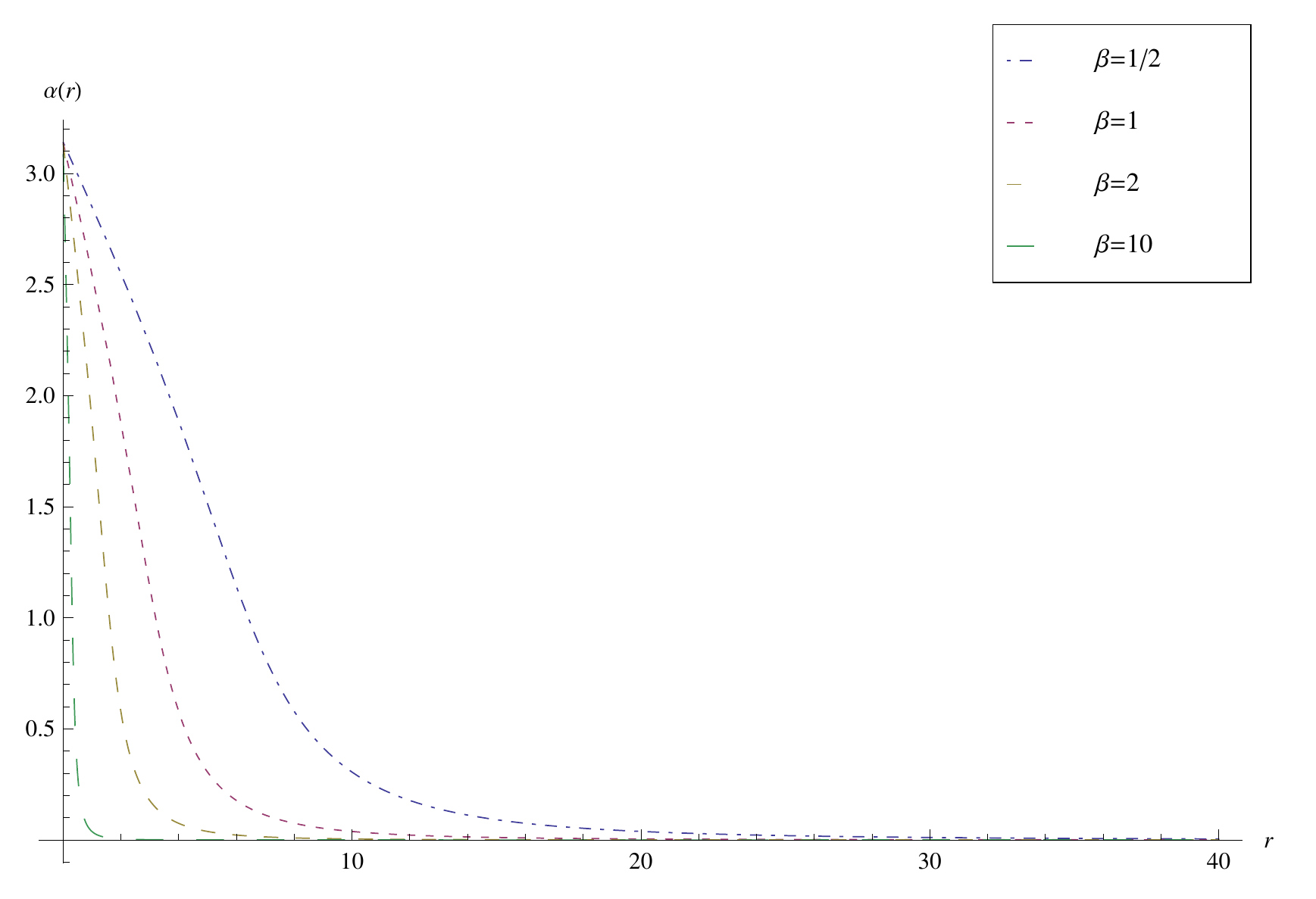}
\caption {$O(4)$ DBI Textures (in $N=4$) with various DBI coupling values.}
\label{hopfbeta}
\end{figure}

The size of solutions for each codimension $N$ is characterized by the parameter $\beta$. This can be shown by variational principle; the soliton gains its natural scale due to the balance of the static energy, {\it i.e.}, its scale is the size ($\ell$) that minimizes the energy. In our theory the energy is roughly given by~\cite{dbipaper}
\begin{equation}
E\sim\ell^{N} \beta^{2}\left(\sqrt{1-\frac{1}{\beta^{2}\ \ell^{2}}}-1\right),
\end{equation}
so the defects size is roughly 
\begin{equation}
\label{sizes}
\ell\sim\frac{(N-1)}{\sqrt{N(N-2)}}\ \beta^{-1}.
\end{equation}
Codimension-$3$ solutions with various $\beta$ are shown in Fig.\ref{hopfbeta}, where it reveals the property of (\ref{sizes}). As $\beta$ increases the scale of defect becomes smaller. We can extrapolate this result to conclude that at $\beta\rightarrow\infty$ the defect is infinitely thin. This result agrees with Derrick's theorem, since for $\beta\rightarrow\infty$ our Lagrangian (\ref{eq:chiraldbisigma}) reduces to the ordinary $O(4)$ textures in codimension three, which possesses no static solution other than the trivial vacuum.
 
In this letter we address some 'leftover' from our previous work on extended DBI solitons in arbitrary codimensions~\cite{dbipaper}. We show here that without the need of upgrading the ansatz symmetry into $O(N)$ we can still find solitonic solutions. These solutions fall into the category of $\pi_{N}(S^{3})$ classes. The simplest nontrivial case, $N=4$, is the direct generalization of the well-known Hopf fibration. 
 
To achieve stability, we consider scalar field theory with Dirac-Born-Infeld (DBI) Lagrangian. In principle there is no obstruction to study solitonic solutions in other models with noncanonical kinetic term(s), such as those proposed in~\cite{noncanon, avelino}. However, as we argued in~\cite{dbipaper}, the DBI form is one of the most economical way to generate higher order kinetic terms by means of Taylor expansion. Furthermore this form can be motivated by some string theory models, as in~\cite{sarangi}Thus in this sense the DBI kinetic term is the most natural example of class of field theories with noncanonical kinetic terms.
 
Although we found numerical solutions within the spherically symmetric ansatz we have not made attempt to justify that this ansatz is unique, nor do we try to check the stability of the solutions~\footnote{In~\cite{faddeev} the only stable solution with $\pi_{3}(S^{2})$ Hopf class possesses toroidal, instead of spherical, symmetry.}. It is interesting to prove or disprove the stability of spherically symmetric ansatz. We hope to address this issue in the future work.

We thank Dionisio Bazeia, Jose Blanco-Pillado, Nuno Romao, and Tanmay Vachaspati for enlightening discussions. This work was begun while the author was attending the Spring School on Superstring Theories and Related Topics at the Abdus Salam ICTP in Trieste, Italy. We thank the institution for their hospitality.


\end{document}